\documentclass[conference]{IEEEtran}

\usepackage{cite}
\usepackage{amsmath,amssymb,amsfonts,color,colortbl}
\usepackage[font=footnotesize]{caption}
\usepackage{algorithmic}
\usepackage{multirow,makecell,textgreek}
\usepackage{balance}
\usepackage{float,flushend}
\usepackage{graphicx}\usepackage{subcaption}
\usepackage{tcolorbox}

\usepackage[export]{adjustbox}
\def\BibTeX{{\rm B\kern-.05em{\sc i\kern-.025em b}\kern-.08em
		T\kern-.1667em\lower.7ex\hbox{E}\kern-.125emX}}

\ifCLASSINFOpdf
\else
\fi
\begin{document}
\title{Evaluation of Switching Technologies for Reflective and Transmissive RISs at Sub-THz Frequencies}
\author{\IEEEauthorblockN{  
		Sofia I. In\'{a}cio$^1$, Yihan Ma$^2$, Qi Luo$^2$, Luca Lucci$^3$, Awanish Kumar$^3$, Jos\'{e} Luis Gonzalez Jimenez$^3$, Bruno Reig$^3$,\\Alexandre Siligaris$^3$, Denis Mercier$^3$, Jonas Deuermeier$^4$, Asal Kiazadeh$^4$, Verónica Lain-Rubio$^5$, Oleg Cojocari$^5$,\\Tung D. Phan$^6$,Ping Jack Soh$^6$, S\'{e}rgio Matos$^7$, George C. Alexandropoulos$^8$, Lu\'{i}s M. Pessoa$^1$, Antonio Clemente$^3$}
		\\
		\IEEEauthorblockA{$^1$INESC TEC, Faculdade de Engenharia, Universidade do Porto, Portugal, $^2$University of Hertfordshire, UK, $^3$Univ Grenoble\\Alpes, CEA, Leti, France, $^4$NOVA University Lisbon, Portugal, $^5$ACST, Germany, $^6$University of Oulu, Finland, $^7$Instituto\\de Telecomunicacoes, University Institute of Lisbon, Portugal, $^8$National and Kapodistrian University of Athens, Greece}
}

\maketitle

\begin{abstract}
For the upcoming 6G wireless networks, reconfigurable intelligent surfaces are an essential technology, enabling dynamic beamforming and signal manipulation in both reflective and transmissive modes. It is expected to utilize frequency bands in the millimeter-wave and THz, which presents unique opportunities but also significant challenges. The selection of switching technologies that can support high-frequency operation with minimal loss and high efficiency is particularly complex. In this work, we demonstrate the potential of advanced components such as Schottky diodes, memristor switches, liquid metal-based switches, phase change materials, and RF-SOI technology in RIS designs as an alternative to overcome limitations inherent in traditional technologies in D-band (110–170 GHz).
\end{abstract}

\begin{IEEEkeywords}
	Reconfigurable intelligent surfaces, switching technologies, sub-THz, D-band, unit-cell design.
\end{IEEEkeywords}

\IEEEpeerreviewmaketitle

\section{Introduction}

In recent years, reconfigurable intelligent surfaces (RISs) have attracted notable interest due to their potential to revolutionize wireless communication systems~\cite{huang2019reconfigurable}. Due to their ability to dynamically manipulate electromagnetic waves, RISs grant to enhance signal quality, improve energy efficiency, and empower control over wireless propagation environments~\cite{APK2023}. In this framework, sub-terahertz (sub-THz) frequencies are gaining significant prominence as they provide abundant spectrum resources to satisfy the growing needs for increased data rates and capacity in future networks~\cite{ETSI_THz}. The D-band, which encompasses frequencies from 110~GHz to 170~GHz, is well-suited for addressing this requirement, serving as an interesting testing band for innovative technologies and architectures, including RISs~\cite{AJG2024}. Although D-band advantages include relatively compact hardware dimensions due to its shorter wavelength, enabling the design of highly integrated and densely packed systems, operating at these frequencies comes with unique challenges. These challenges include higher free-space path loss, increased susceptibility to atmospheric absorption, and the necessity for advanced materials and components capable of efficient signal manipulation~\cite{TERRAMETA_Eucap2024}.

RISs can be distinguished by their ability to redirect or let electromagnetic waves pass through, in reflective (R-RIS) and transmissive (T-RIS), respectively~\cite{BAL2024}. A key factor in developing efficient RISs is the selection of suitable switching technologies, which are essential for the precise modulation of electromagnetic waves at the unit-cell level~\cite{RIS_THz_terrameta}. As the fundamental components, unit-cells are designed to allow the control of the phase, amplitude, or polarization of incident signals. Unlike traditional phased-arrays, which rely on phase shifters and power amplifiers, RISs are typically passive radiative structures. They typically integrate components such as RF-MEMS, PIN diodes, varactors, or liquid crystals to control the local surface phase shift and impedance characteristics~\cite{HMIMO_survey}. Based on the technology used in the unit-cell, RISs can be classified as active, semi-passive, or passive~\cite{BAL2024}. Active RISs amplify the reflected signal using active elements in the unit-cells, semi-passive ones use active reception unit for signal absorption and processing~\cite{ASA2024}, while passive RISs use low-loss reactive components to implement either continuous or quantized phase shifts, resulting in energy-efficient devices.

Switching elements, such as PIN diodes, commonly used in RIS for 5G communications, prove to be impractical at THz and sub-THz due to their cutoff frequency and increased loss in this range~\cite{venkatesh2020high}. Consequently, a variety of tuning mechanisms have been investigated recently. This work aims to analyze, design and validate diverse reconfigurable technologies for RIS operating at D-band. These include electronic approaches (Schottky diodes, memristor switches and radio frequency silicon on insulator (RF-SOI)), phase-change materials (PCM) and liquid metal (LM). These switching technologies have several advantages over those commonly used in RIS. Schottky diodes and Memristor enable faster switching capabilities~\cite{elsaid2024non, yang2022terahertz}, allowing dynamic beamforming and real-time signal manipulation. Schottky diodes and RF-SOI technology feature lower insertion losses than conventional PIN diodes and MEMS switches, which enables better overall efficiency~\cite{RIS_THz_terrameta}. Memristors, with their non-volatile characteristics and the efficient phase transitions of PCMs, allow a reduction in energy consumption~\cite{dragoman2017phased}. RF-SOI technology facilitates greater integration density and scalability, allowing for more compact and complex designs~\cite{RIS_THz_terrameta}. These components and technologies present a promising alternative for D-band RIS applications.

The paper is organized as follows: Section~\ref{switching} describes the switching technologies at D-band studied in this work. Section~\ref{R-RIS} details the R-RIS architectures developed for the D-band, including designs based on Schottky diodes, memristor switches, and LM materials. Section~\ref{T-RIS} focuses on the programmable T-RIS solutions at the D-band, leveraging PCMs and CMOS-based switches for enhanced performance and control. The conclusions are summarized in Section~\ref{conclusion}.

\section{Switching Technologies at D-band}~\label{switching}

The RIS unit-cell design follows the local periodicity principle~\cite{pozar1997design}, where the scattering properties of each sub-wavelength element (typically below $\lambda/2$ to avoid grating lobes) resemble those of a periodic structure. The RIS response arises from the superposition of these scattering elements. Reconfigurable phase control at the unit-cell level enables dynamic shaping of both near- and far-field radiation patterns of the RIS.

There are four main methods for phase control:

\begin{itemize}
	\item[-] Parasitic tunable elements: Adjusting resonance via discrete or continuous elements, though bandwidth is limited, and insertion losses are higher.
	\item[-] Transmission line switching: Switching current paths within the cell, offering higher bandwidth but increasing circuit complexity.
	\item[-] Electronic rotation: Using multiple switches to modify current paths, emulating physical rotation with lower losses.
	\item[-] Tunable shape: Modifying metallization via microfluidic actuation, allowing continuous phase control but with slower response times.
\end{itemize}

\noindent Besides these, tuning techniques based on phase-change materials, graphene and optical tuning provide viable alternatives, particularly for high-frequency and energy-efficient applications~\cite{liu2021reconfigurable}. Each approach has trade-offs, with the selection depending on application needs, complexity, bandwidth, and response speed. This section describes the synthesis, design, simulation, and characterisation of R-RISs and T-RISs using different switching methods.

\subsection{Reflective RIS} \label{R-RIS}

\subsubsection{Schottky diodes}
Traditionally, high losses have constrained the use of diodes above mmWave frequencies. However, there is growing interest in advanced diode technologies, particularly Schottky-based designs, which are engineered to minimize parasitic effects that hinder high-frequency performance~\cite{vassos2021air, shen2025sub}. The Schottky-based devices achieve switching speeds below a picosecond, with their ultimate limit determined by the biasing circuitry rather than the tunable material itself.

\begin{figure}[!h]
	\centering
	\subfloat[]{{\includegraphics[scale=0.35]{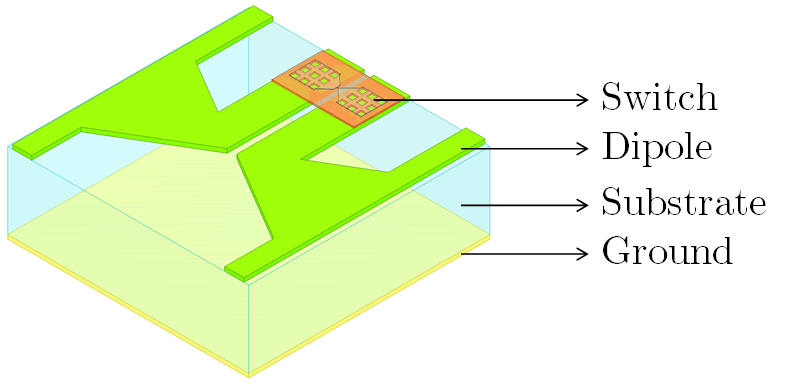} }}
	\\
	\subfloat[]{{\includegraphics[scale=0.24]{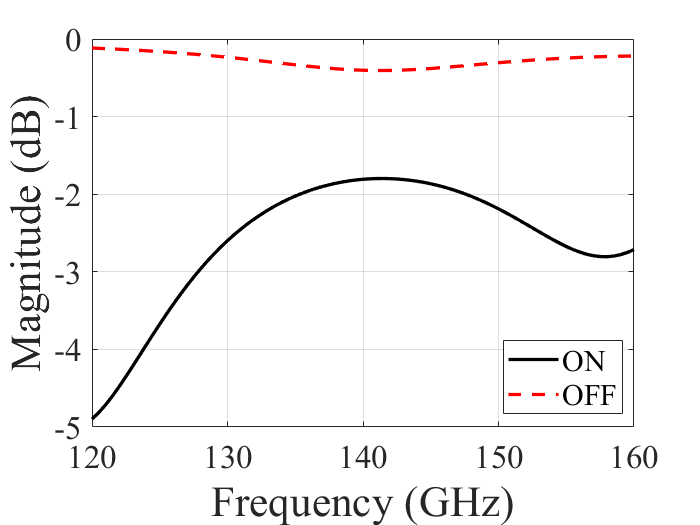} }}
	\subfloat[]{{\includegraphics[scale=0.24]{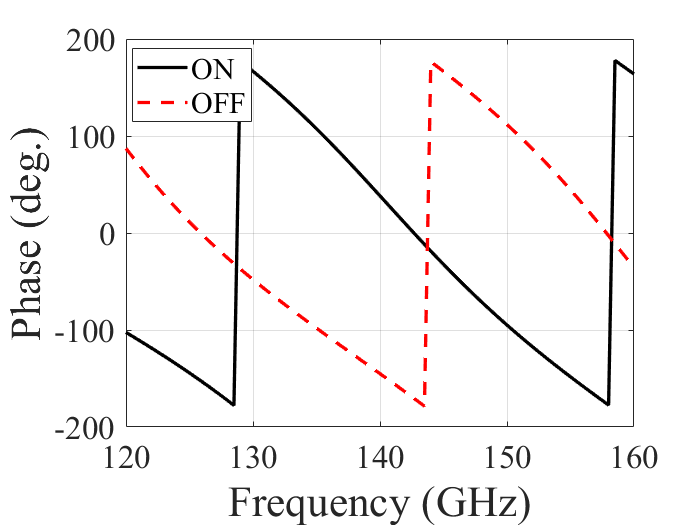} }}
	\caption{(a) Configuration of the designed element and simulated (b) magnitude and (c) phase of the reflection coefficient.}
	\label{fig:uc_schottky}
\end{figure}

Leveraging these advantages, the designed unit-cell comprises planar tightly coupled dipoles (PTCDs). A bowtie dipole is printed on the substrate, and a Schottky diode is integrated between the two patches to control the reflection states of the unit-cell, switching between the ``ON" and ``OFF" states. Fig.~\ref{fig:uc_schottky}(a) shows the simple-to-fabricate and easy-to-integrate unit-cell configuration to achieve active R-RIS. Eliminating the need for drilled holes on the unit-cell enhances the overall mechanical durability and reliability, making the unit-cell particularly attractive for applications where cost-effectiveness and reliability are key considerations. The substrate used is the Astra MT77 dielectric with a standard thickness of 0.3175 mm. Fig.~\ref{fig:uc_schottky}(b) and Fig.~\ref{fig:uc_schottky}(c) show the reflection amplitude and reflection phase of the unit-cell at different states.

\begin{figure}[!h]
	\centering
	\includegraphics[scale=0.3]{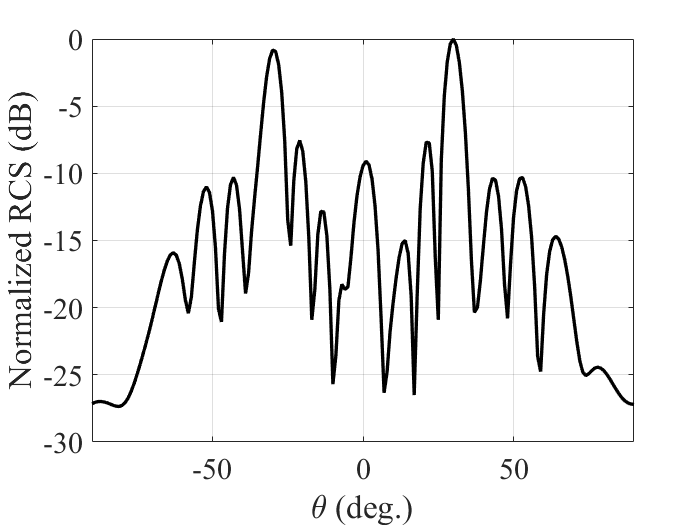}
	\caption{2D RCS of the metasurface scanning at an angle of 30$^\circ$, in the cut-plane $\varphi = 0$.}
	\label{fig:ris_schottky}
\end{figure}

The performance of a $20\times20$~mm$^2$ array aperture, comprising $20\times20$ elements, is validated using the Ansys Electronic HFSS software. This aperture is designed to reflect a normally incident wave at an angle of $30^\circ$. The simulated Radar Cross Section (RCS) result is presented in Fig.~\ref{fig:ris_schottky}. As demonstrated, despite the phase errors introduced by the 1-bit phase quantization, the beam steering error remains minimal.

\subsubsection{Memristor switches}

Memristors provide a non-volatile switching mechanism, eliminating the need for continuous voltage application and thereby enhancing energy efficiency. Their compact size (1–5~$\mu$m), compatibility with printing techniques, high switching speed (nanosecond scale), and low actuation voltage (1–2~V) facilitate streamlined fabrication and integration. These attributes position memristors as a promising candidate for fully printed RIS manufacturing, potentially obviating the need for soldering and conventional packaging processes.

\begin{figure}[!h]
	\centering
	\subfloat[]{{\includegraphics[scale=0.35]{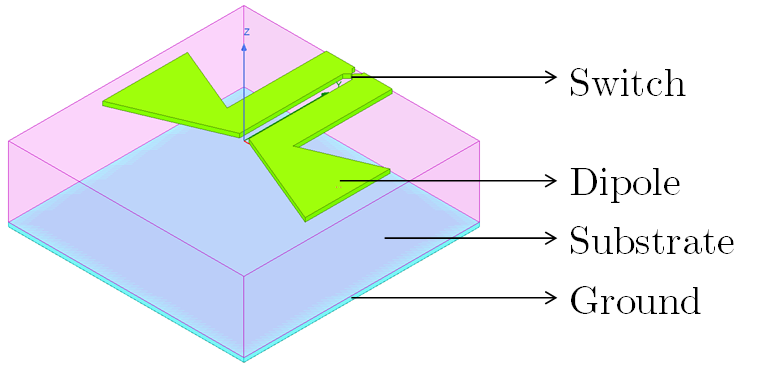} }}
	\\
	\subfloat[]{{\includegraphics[scale=0.24]{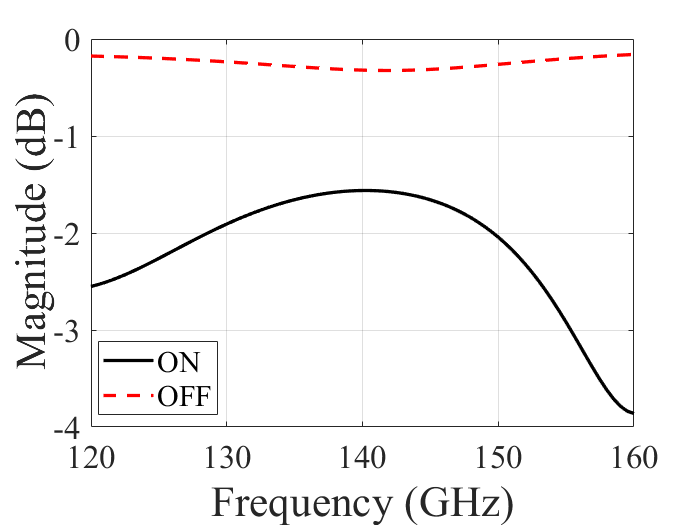} }}
	\subfloat[]{{\includegraphics[scale=0.24]{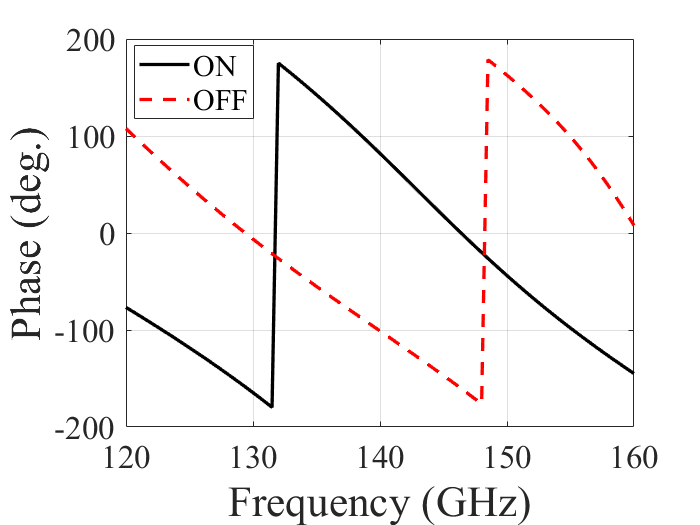} }}
	\caption{(a) Configuration of the designed element and simulated (b) magnitude and (c) phase of the reflection coefficient.}
	\label{fig:uc_memristor}
\end{figure}

Leveraging these advantages, a memristor was used to design a unit-cell with a structure similar to the previous one. The substrate is made of dielectric fused silica with a standard thickness of 0.3~mm. Fig.~\ref{fig:uc_memristor}(a) illustrates the configuration of the designed planar PTCDs. A bowtie dipole is printed on the substrate, and a memristor switch is integrated between the two patches to control the reflection states of the unit-cell, allowing it to switch between the ``ON'' and ``OFF'' states. Fig.~\ref{fig:uc_memristor}(b) and Fig.~\ref{fig:uc_memristor}(c) display the reflection amplitude and phase of the cell in different states, respectively.

\begin{figure}[!h]
	\centering
	\includegraphics[scale=0.3]{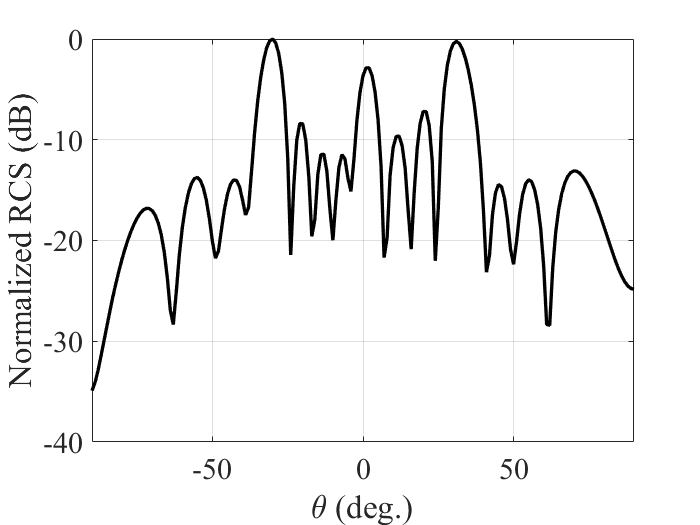}
	\caption{2D RCS of the metasurface scanning at an angle of 30$^\circ$, in the cut-plane $\varphi = 0$.}
	\label{fig:ris_memristor}
\end{figure}

The performance of a $20 \times20$~mm$^2$ array aperture, comprising $20\times20$ elements, was validated using the Ansys HFSS. Fig.~\ref{fig:ris_memristor} shows the normalized scattering pattern with a steering angle of 30$^\circ$. In contrast to the Schottky-diode-based metasurface array, the memristor-based metasurface array exhibits slightly higher sidelobes while remaining within an acceptable range.

\subsubsection{Liquid metal-based reconfigurable beamsplitter}

Most liquid metals have very good electrical conductivity, which is much higher than water and ten times below copper. Among liquid metals, gallium and its alloys are non-toxic, and this is the main reason of its choice for reconfigurable electromagnetic application.

\begin{figure}[!h]
	\centering
	\includegraphics[width=0.75\linewidth]{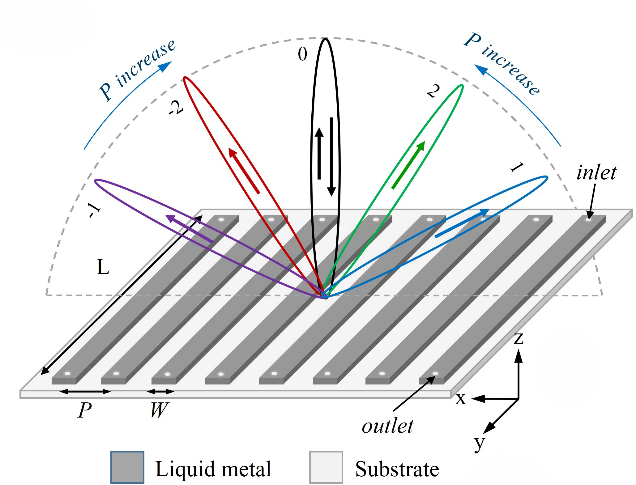}
	\caption{Concept of liquid-metal-based R-RIS working as a reconfigurable beam splitter.}
	\label{fig:concept_lm}
\end{figure}

\begin{figure}[!h]
	\centering
	\includegraphics[width=0.85\linewidth]{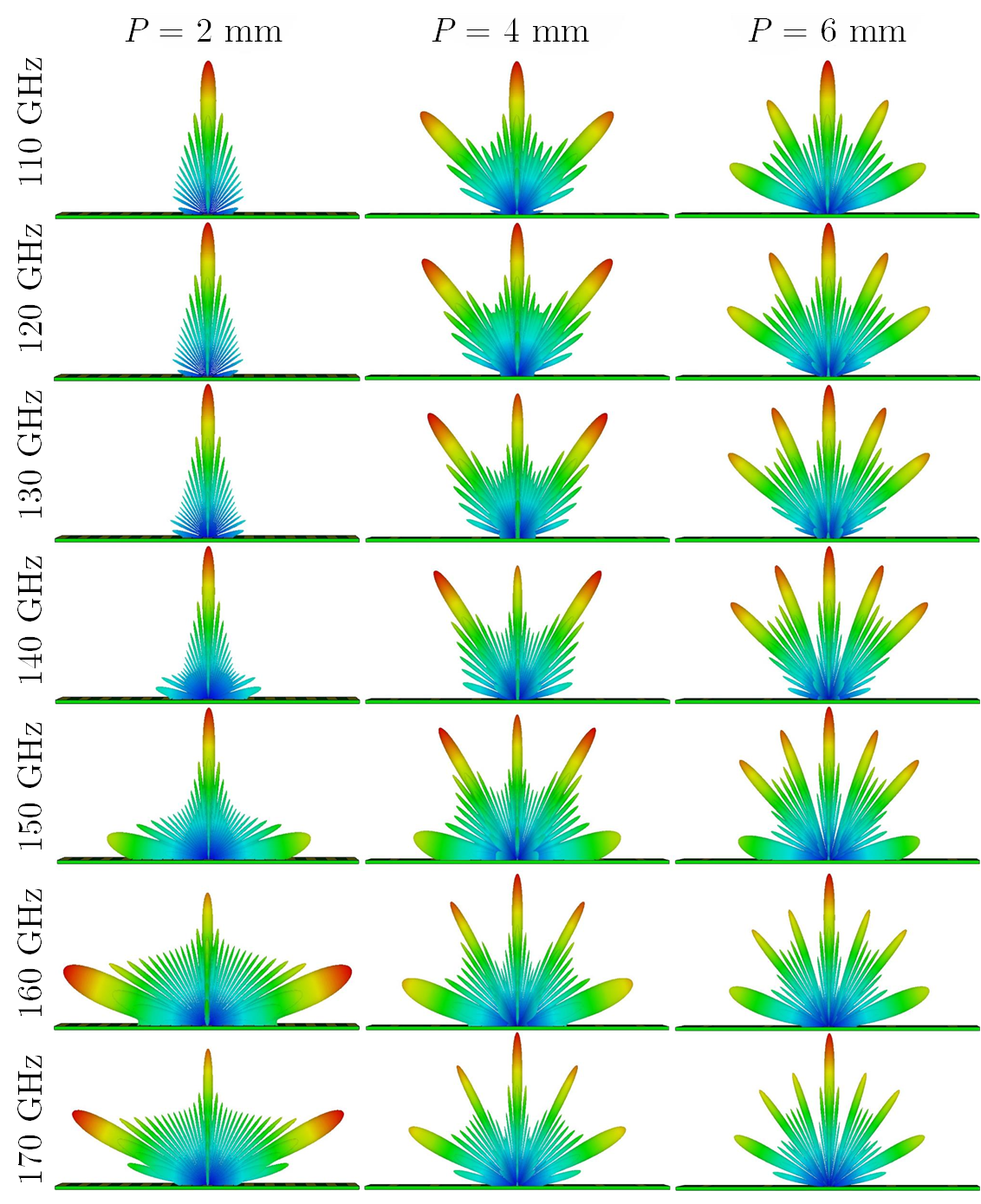}
	\caption{Simulation results of the proposed LM structure at various frequencies within the D-band, for $P = 2$~mm, $P = 4$~mm and $P=6$~mm.}
	\label{fig:sigmulation_lm}
\end{figure}

A reconfigurable structure based on liquid metal (EGaIn) was proposed and designed. This structure operates based on the grating structure principle, consisting of a periodic array of metallic strips on a substrate backed by a ground plane. As shown in Fig.~\ref{fig:concept_lm}, each liquid channel has an in/outlet which allows individual pumps of liquid metal in/out to/from each channel. When each channel is fully filled ($L = L_{max}$) or emptied ($L = 0$), the spacing between the channels, $P$, can be adjusted, and therefore reconfigures the deflected beams. The structure was then fabricated using a low-loss polyimide ($\varepsilon = 3.2$, $\tan \delta = 0.005$) with a thickness of $t_{sub} = 1000~\mu$m. A continuous metal layer to limit the transmitting wave components backs the substrate. The depth of each microchannel is $t_{ch} = 500~\mu$m, and they are covered by a thin Kapton tape layer (thickness of $10~\mu$m) to keep the liquid metal in the microchannels. Fig~\ref{fig:sigmulation_lm} presents the simulation results of the proposed structure at various frequencies within the D-band, as well as under different structural configurations.

\begin{figure}[!h]
	\centering
	\subfloat[]{{\includegraphics[width=0.74\linewidth]{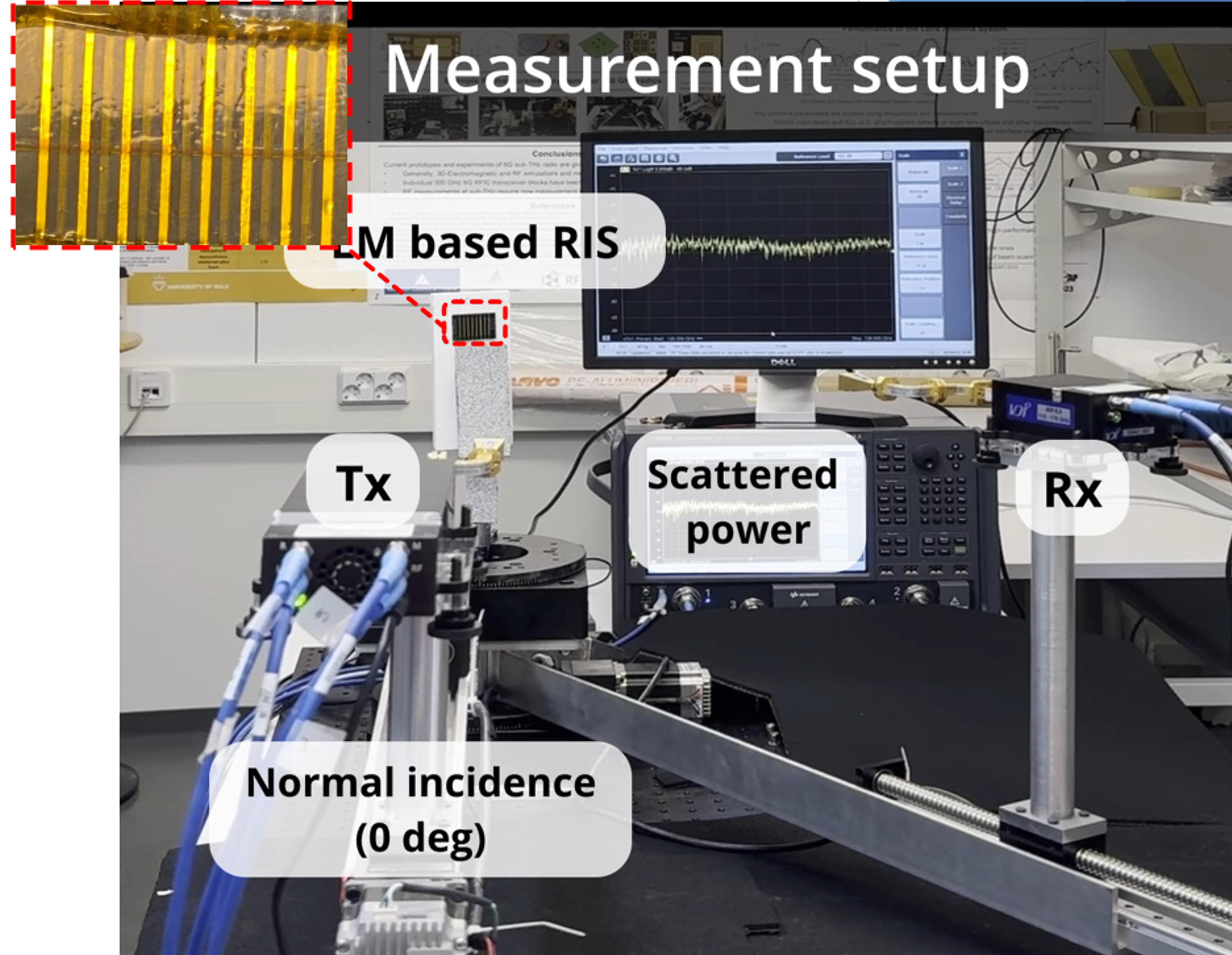} }}
	\qquad
	\subfloat[]{{\includegraphics[scale = 0.33]{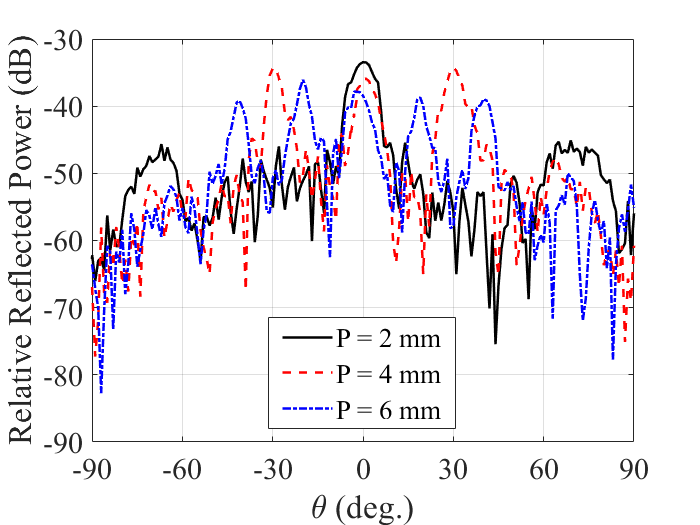} }}
	\caption{(a) Measurement setup and (b) results at 150~GHz for different periodicity: $P = 2$~mm, $P = 4$~mm and $P = 6$~mm.}
	\label{fig:lm}
\end{figure}

Fig.~\ref{fig:lm}(a) illustrates the measurement setup for the liquid-metal-based RIS. The transmit (Tx) horn antenna is positioned 50~cm away from the device-under-test (DUT) and fixed at normal direction, while the receive (Rx) antenna was rotated in a single plane, at the same distance, to capture the scattering pattern from the DUT. Fig.~\ref{fig:lm}(b) presents the measured radiation patterns of the fabricated LM-based RIS, which aligns well with the simulation results. The DUT acts as a reconfigurable beam splitter, as the insertion and removal of the liquid metal into/from the microchannels modulates the distance between the subsequent channels. This modulation enables control of the scattered beam, allowing the RIS to reflect in various configurations: When $P = 2$ mm (which equals $\lambda$ at 150~GHz), only one propagating mode is active ($n = 0$), resulting in the reflection of the incident beam in the normal direction. On the other hand, when $P = 4$ mm and 6 mm, the number of propagating modes increases to three and five, respectively. As designed, the single incident beam will be split into three and five beams, respectively.

\subsection{Transmissive RIS} \label{T-RIS}

\subsubsection{Phase change material}

Phase change material (PCM) based switches have emerged as a promising candidate for high-frequency reconfigurable systems. These switches exhibit a reversible transition between amorphous and crystalline states, allowing for tunable electromagnetic properties along with non-volatile switching capabilities~\cite{welnic2008reversible}. This intrinsic feature enables PCM-based switches to maintain a specific state without requiring a continuous power supply, thereby significantly enhancing energy efficiency in RIS applications.

\begin{figure}[!h]
	\centering
	\includegraphics[width=0.86\linewidth]{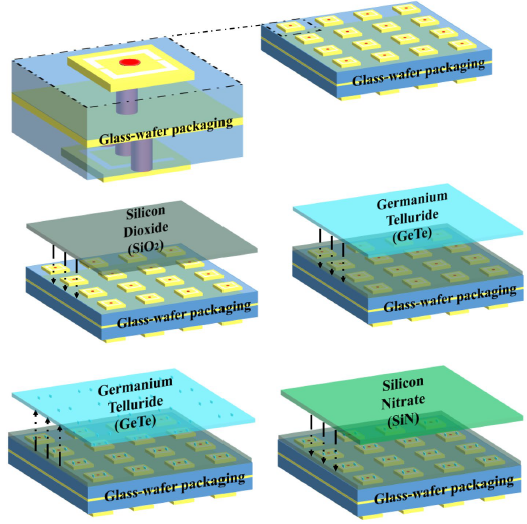}
	\caption{Concept of the PCM-based T-RIS unit-cell and its proposed fabrication process in clean room.}
	\label{fig:uc_pcm}
\end{figure}

The proposed unit-cell (Fig.~\ref{fig:uc_pcm}) consists of a wafer-like topology with three metal layers: a passive patch antenna, a ground plane, and an active patch antenna with integrated PCM switches. Two quartz glass dielectrics, with $\varepsilon_r = 3.78$ and $\tan \delta = 0.0001$, separate these layers. Optimized for the D-Band, the periodicity is set to $\lambda/2$ at 140~GHz. The active rectangular patch, made from Aluminium–copper (AlCu), includes an O-slot and two germanium tellurium (GeTe) switches that provide symmetrical scattering behavior with a $180^\circ$ phase shift by alternating between ``ON'' (crystalline) and ``OFF'' (amorphous) states, resulting in a 1-bit unit-cell design. The transmitting patch connects to a passive receiving patch through a metalized via hole, while the passive patch, which is rectangular and loaded with a U-slot, connects to the ground plane via two vertical vias. 

\begin{figure}[!h]
	\centering
	\subfloat[]{{\includegraphics[scale=0.24]{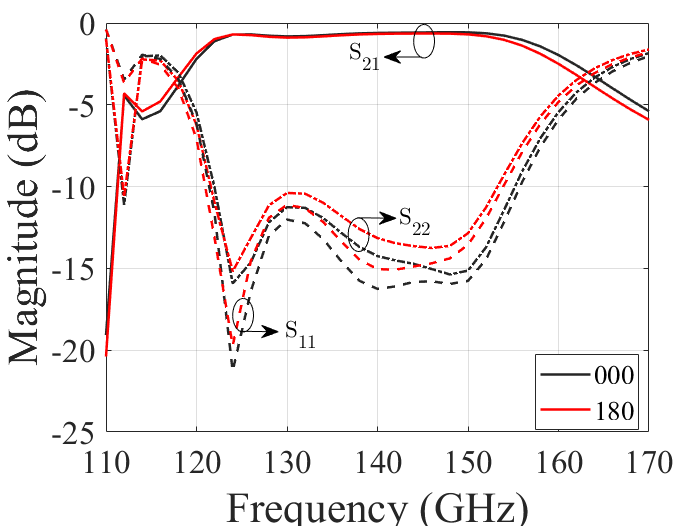} }}
	\subfloat[]{{\includegraphics[scale=0.24]{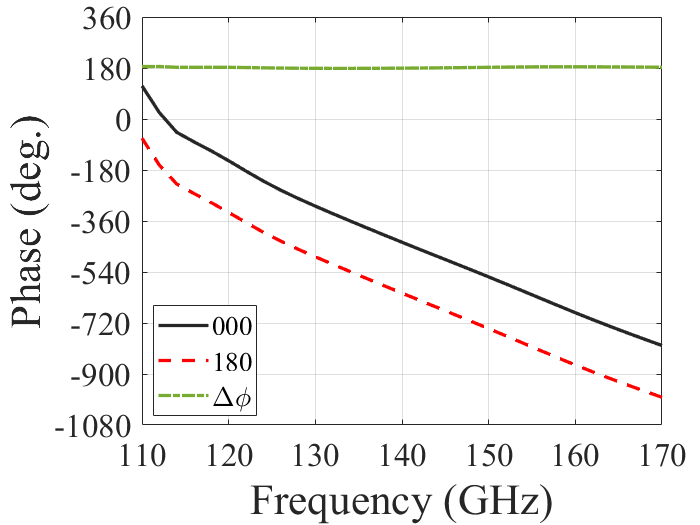} }}
	\caption{Simulated scattering parameters of the proposed reconfigurable unit-cell with PCM based switches. (a) Magnitude of the transmission and reflection coefficients and (b) phase of the transmission coefficients for the phase states 000 and 180. The achieved phase difference ($\Delta \phi$) is also reported in (b).}
	\label{fig:uc_pcm_s}%
\end{figure}

The proposed unit-cell performance was studied using full-wave simulations (using Ansys HFSS) with periodic boundary conditions and Floquet ports. The simulations demonstrate the reconfigurability of the unit-cell with GeTe switches in two-phase states (called 000 and 180). Fig.~\ref{fig:uc_pcm_s}(a) shows the S-parameter magnitudes for these states, indicating a 0.69~dB insertion loss at 140~GHz and losses below 1.5~dB over a 27\% bandwidth. Fig.~\ref{fig:uc_pcm_s}(b) displays the phases of $S_{21}$ for both states, revealing a consistent $180^\circ$ phase difference and low insertion losses over a relatively large frequency range.

\begin{figure}[!h]
	\centering
	\includegraphics[width=1\linewidth]{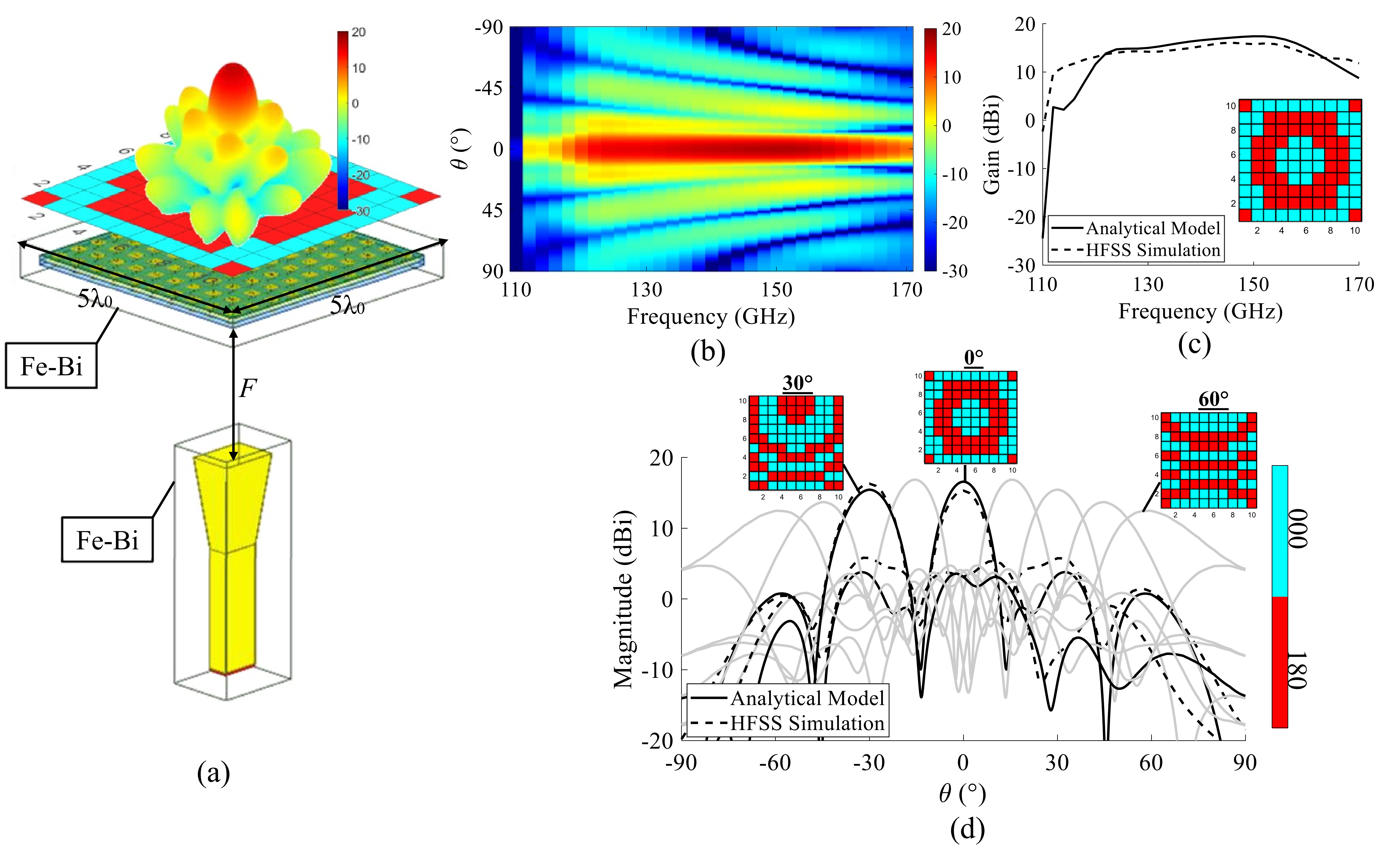}
	\caption{(a) The T-RIS simulation setup in HFSS. (b) The gain in cut-plane $\varphi = 0$ in terms of $\theta$ and the frequency calculated in Matlab. (c) The comparison of gain variation in the frequency band for the broadside transmission in the cut-plane $\varphi = 0$ between HFSS and MATLAB. (d) The comparison of the radiation pattern for several transmissions between HFSS and MATLAB.}
	\label{fig:TRIS_pcm}
\end{figure}

This unit-cell has been used to design a 100-element square T-RIS, which has been synthesized using an in-house numerical model (MATLAB analytical model) and validated using the commercial full-wave software Ansys Electronic HFSS \cite{Clemente24}. The proposed simulation setup, the numerical performances of the T-RIS and the comparison between the different simulation tools are presented in Fig.~\ref{fig:TRIS_pcm}. The maximum broadside gain at 140 GHz exported from HFSS is of 15.8 dBi, which is slightly lower than the gain calculated in MATLAB (16.6 dBi).

\subsubsection{45 nm RF-SOI technology}

RF-SOI (Radio Frequency Silicon-on-Insulator) switches are a key enabler for high-performance RIS, especially above mmWave frequencies. In contrast to traditional CMOS-based switches, RF-SOI technology features a buried oxide (BOX) layer, which minimizes substrate losses and enhances isolation and insertion loss~\cite{Rebeiz12}. RF-SOI switches demonstrate significantly lower power consumption when compared to PIN diodes. Widely adopted in industry, RF-SOI switches have been successfully used up to 220 GHz with excellent isolation and insertion losses.

\begin{figure}[!h]
	\centering
	\subfloat[]{{\includegraphics[scale=0.3]{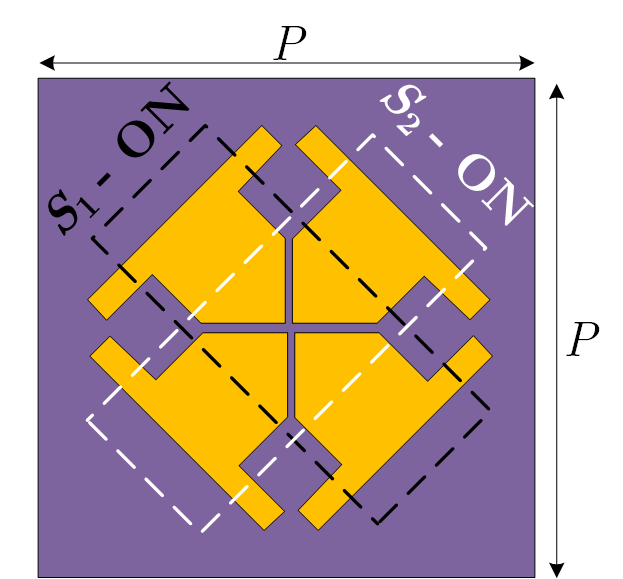} }}
	\subfloat[]{{\includegraphics[scale=0.26]{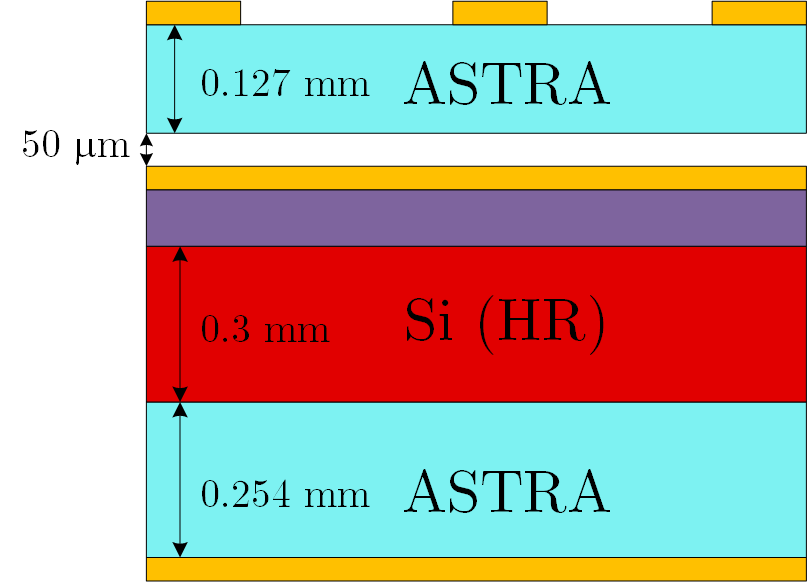} }}
	\caption{Front and side view of the design T-RIS unit-cell ($P = 0.78~$mm$~= 0.37\lambda_L$).}
	\label{fig:uc_cmos}
\end{figure}

The unit-cell is designed with a Fabry-Perot transmissive resonator, where the transmission phase depends on the shape and orientation of the rotators~\cite{Transmitarray_THz_2023}. When two identical rotators are oriented oppositely, their transmission phases are reversed. This design incorporates two diagonally arranged I-shaped rotators, each connected to a switch (S$_1$ or S$_2$), ensuring that only one switch is on at any time, allowing for 1-bit phase quantization. The unit-cell uses CMOS hybrid-PCB packaging technology and consists of seven vertically stacked layers. The top and bottom layers are orthogonal polarisers printed on Astra MT-77 substrate, while the EM rotators and CMOS switches are printed on a 3~$\mu$m SiO$_2$ coated high-resistivity (HR) silicon substrate, sandwiched between the Astra MT-77 layers. In the unit-cell optimization, a 50~$\mu$m air gap has been considered between the top Astra MT-77 substrate and HR-silicon to take into account the space required to integrate the RFIC on the dielectric substrate. The CMOS switch consists practically of the combination of parallel RC circuits. In particular, the value of $C$ and $R$ of CMOS switches are dependent on CMOS dimensions, channel mobility, fabrication process, and substrate doping. The approximate equivalent circuit model of CMOS switches and their value in both ``ON'' and ``OFF'' cases are $R_{\text{on}} = 6.13~\Omega$, $C_{\text{on}} = 18.5~$fF, $R_{\text{off}} = 4300~\Omega$, and $C_{\text{off}} = 19.0~$fF, respectively. Fig.~\ref{fig:uc_cmos} shows the front and side views of the unit-cell with its geometric parameters.

\begin{figure}[!h]
	\centering
	\subfloat[]{{\includegraphics[scale=0.24]{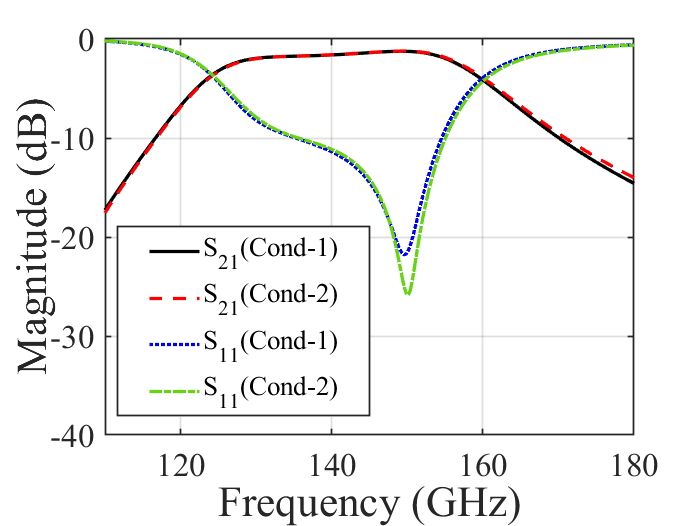} }}
	\subfloat[]{{\includegraphics[scale=0.24]{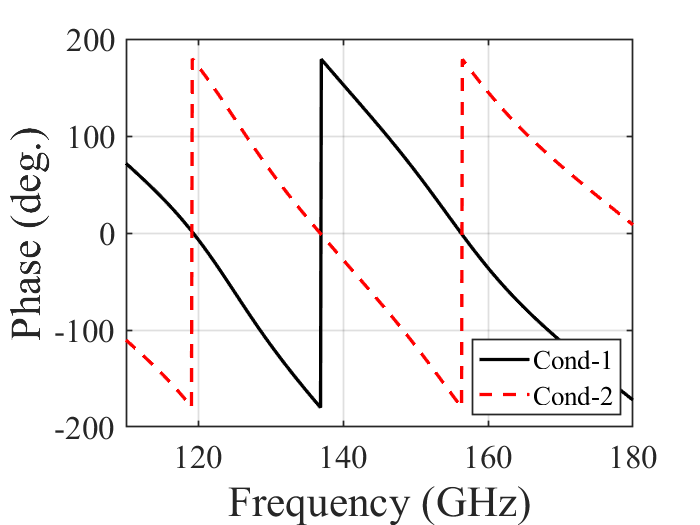} }}
	\caption{(a) Unit-cell with an air-gap of 50~$\mu$m (b) S-parameter over 110-180~GHz.}
	\label{fig:uc_cmos_2}
\end{figure}

The proposed unit-cell has been simulated using CST microwave studio software with a periodic boundary condition in the XY direction and free-space boundaries. The magnitude and phase of the simulated transmission coefficient are illustrated in Fig.~\ref{fig:uc_cmos_2}(a) and Fig.~\ref{fig:uc_cmos_2}(b), respectively. The designed unit-cell exhibits nearly identical transmission magnitudes but opposite transmission phases for two conditions: ($S_1$-ON, $S_2$-OFF) and ($S_1$-OFF, $S_2$-ON). Additionally, Fig.~\ref{fig:uc_cmos_2}(a) indicates a wide transmission bandwidth of 30~GHz with less than 2~dB return loss over 121 GHz-158 GHz with absolute bandwidth of 37~GHz.

\section{Conclusion} \label{conclusion}
In this paper, we presented several innovative switching technologies for RIS, encompassing both reflective and transmissive designs, that operate within the D-band. We provided both simulation and experimental validation of various RIS architectures, including passive and programmable R-RIS, utilizing various technologies such as Schottky diodes, memristors, and liquid metal.

The progress made in T-RIS technologies is similarly commendable, employing phase-change materials and CMOS switches, further expanding the technological frontier of RIS capabilities. Each of these technologies offers distinct advantages, positioning them as strong candidates for overcoming the limitations of conventional approaches.

This work has demonstrated significant progress in switching technologies for RIS at sub-THz frequencies. The future integration of these final RIS architectures will, to the best of our knowledge, mark a significant advancement in the field of sub-THz applications.

\section*{Acknowledgment}

This work was supported by the SNS JU project TERRAMETA under the European Union's Horizon Europe research and innovation programme under Grant Agreement No 101097101, including top-up funding by UK Research and Innovation (UKRI) under the UK government's Horizon Europe funding guarantee. This work was also supported by FCT/MCTES under grant 2020.04948.BD. A part of the work carried out at CEA Leti has been as well partially supported by the France 2030 recovery plan as part of the ``PEPR'' program via the projects SYSTERA and FUNTERA.

\bibliographystyle{IEEEtran}

\bibliography{references}

\end{document}